\documentclass[12pt]{article}

\usepackage{latexsym}

\begin{document}

\title{Poisson Algebra of Diffeomorphism Generators in a Spacetime Containing a Bifurcation}
\author{A. Giacomini\thanks{e-mail: giacomin@science.unitn.it}\\
Dipartimento di Fisica, Universit\`a di Trento\\
and Istituto Nazionale di Fisica Nucleare,\\
Gruppo Collegato di Trento, Italia}
\date{}
\maketitle

\begin{abstract}
In this article we will analyze the possibility of a nontrivial central extension of the Poisson algebra of the diffeomorphisms generators 
which respect certain boundary conditions on the black hole bifurcation. The origin of a possible central extension in the algebra is due to the
existence of boundary terms in the canonical generators, which are necessary to make them differentiable. The existence of such 
boundary terms depends on the exact boundary conditions that one takes. We will check two possible boundary conditions on the
black hole bifurcation: Fixed metric and fixed surface gravity. In the case of fixed metric of the bifurcation the action
acquires a boundary term but this term is canceled in the Legendre transformation and so absent in the Hamiltonian, and so in this
case the possibility of a central extension is ruled out. 
In the case of fixed surface gravity the boundary term in the action is absent but therefore present in the Hamiltonian.
Also in this case case we will see that there is no central extension, also if there exist boundary terms in the generators.

\end{abstract}

\section{Introduction}

Since the discovery of the Bekenstein-Hawking entropy \cite{hawking&bardeen&carter}, \cite{hawking}, \cite{bekenstein},
\cite{bekenstein2} for black holes given by 
\begin{equation}
S_{BH}  =   \frac{A}{4}
\end{equation} 
a great effort has been done to find a microscopic interpretation of this entropy.\\
As it is well known no complete quantum theory of gravity is available at the moment. There exist of course some specific
models of quantum gravity like superstrings \cite{vafa}, loop quantum gravity \cite{ashtekar} or Sakharov induced gravity \cite{sakharov} which are able to count 
the microscopic degrees of freedom of the entropy for some classes of black holes. Those approaches, which are conceptually very different, always give the 
same result i.e. the Bekenstein-Hawking result. There is therefore a sort of universality principle.\\
It seems therefore that the behavior of the microstates is already set in some way at classical level. 
A classical principle, which is inherited at quantum level is for example a symmetry principle. A symmetry group which is enough
powerful to fix  the density of microstates is the 2-D conformal group where the asymptotic density of states is given by the Cardy formula \cite{cardy}.
The Cardy formula uses only the value of the central charge and of the  $L_0$ generator. A  central charge can already merge at
classical level in the Poisson algebra of canonical generators. \cite{arnold} \\
There are two ways of how to find a 2-D conformal theory with a classical central charge describing the black hole.
One way is by dimensional reduction of the Einstein-Hilbert action obtaining an effective 2-D theory. Making a near horizon approximation
one can  obtain eventually a conformal field theory with a classical central charge using the
radial degree of freedom \cite{solodukhin}, \cite{brustein}, or using as degree of freedom the conformal factor of the 2-metric
as done by A. Giacomini \& N. Pinamonti \cite{giacomini&pinamonti}, and by A. Giacomini
\cite{giacomini2}, \cite{giacomini3}.\\
The other way  inspired by  the work of Brown \& Henneaux,  \cite{brown&henneaux},   
is to construct diffeomorphisms that preserve 
certain fall off conditions of the $r-t$ plane metric of the black hole and then compute the Poisson brackets of the canonical
generators of this diffeomorphisms and check if they form a Virasoro algebra with a nonzero central charge.
Brown \& Henneaux showed in the article cited before that this happens in $AdS_3$ for diffeomorphism generators that preserve the form
of the metric at infinity.\\  
The Poisson bracket of the canonical generators of diffeomorphisms in fact has the form \cite{teitelboim}
\begin{equation}
\left\{  H[\xi ] , H[\eta ]  \right\} = H\left[ [\xi ,\eta ]_{SD}   \right] +K (\xi , \eta )   \; ,    \label{poisson}
\end{equation}
where the bracket $[\, , \, ]_{SD}$ is the so called surface deformation algebra given by
\[
\left[\hat{\xi} , \hat{\eta}   \right] _{SD} ^{\bot} = \hat{\xi} ^a \partial _a \hat{\eta} ^{\bot} - \hat{\eta} ^a \partial _a \hat{\xi} ^{\bot}
\]
\begin{equation}
\left[ \hat{\xi} , \hat{\eta} \right]_{SD} ^a = \hat{\xi} ^b \partial _b \hat{\eta} ^a -\hat{\eta} ^b \partial _b \hat{\xi} ^a 
+ h^{ab}\left( \hat{\xi} ^{\bot} \partial _b \hat{\eta} ^{\bot} - \hat{\eta} ^{\bot} \partial _b \hat{\xi} ^{\bot}  \right)   \;,
\end{equation}
where $h^{ab}$ is the metric of the spacelike hypersurface and we have introduced for $\xi$ the components normal and tangent to the 
foliation
\[
\hat{\xi} ^{\bot} = N \xi ^t 
\]
\begin{equation}
\hat{\xi} ^a = \xi ^a + N^a \xi ^t
\end{equation}    
 The origin of the central extension is due to the existence of arbitrary terms, which do not depend on 
the canonical variables in the generators. In fact the canonical generators in (\ref{poisson}) have the form
\begin{equation}
H[\xi] = \int _{Bulk} \left( \hat{\xi} ^{\bot} \mathcal{H}  + \hat{\xi} ^i \mathcal{H} _i   \right) + J[\xi ] + C(\xi ) \; .     \label{canonicalgenerator}
\end{equation} 
The terms in the bulk integral $\mathcal{H}$ and $\mathcal{H} _i$ are constraints. The Boundary terms $J[\xi ]$ are needed 
to make the generators differentiable and so to make the Poisson brackets well defined . The term $C(\xi )$
is an arbitrary function that does not depend on the canonical variables.
Now it has been shown \cite{brown&henneaux2} that the Poisson bracket of two differentiable generators is also differentiable
and therefore has the correct boundary term. What may happen is that the Poisson bracket does not match the arbitrary term $C(\xi )$
and this is then the origin of the central term $K$.\\
This approach has been used by Strominger \cite{strominger} to compute the entropy of the BTZ black hole \cite{BTZ} using its asymptotic $AdS$ structure
and so the results of Brown \& Henneaux. The problem is that this approach is limited to the BTZ model and its $AdS$ structure at infinity.
An extension to black holes embedded in $AdS_2$ is given in \cite{cadoni1}, \cite{cadoni2}.\\
Using symmetries at infinity one is not able to distinguish a black hole from a star. On the other hand, as explained before
one expects the degrees of freedom responsible for the entropy to live on on or near the horizon. One should therefore study
the diffeomorphisms preserving the near horizon structure instead of spatial infinity.\\     
This approach has been performed in many articles, but there seem to be some technical difficulties \cite{park}, \cite{soloviev}.
In  e.g. \cite{carlip},  \cite{ghosh} the calculation does not work in the case of non-rotating black holes. 
In \cite{carlip2}, \cite{silva}, a covariant formalism  as developed by Wald et al. \cite{wald1} \cite{wald2} \cite{wald3} \cite{wald4} is used 
instead of the original ADM \cite{ADM} approach. A generalization of this approach to Lagrangian of  arbitrary curvature dependence
is found in \cite{pallua2}.\\
In our work we prefer to use 
again the canonical ADM formalism because of its better transparency and it's successful use in the work of Brown \& Henneaux.
A return to the ADM formalism was already tried in \cite{park2}, but the boundary conditions and so the nature of the boundary terms of the canonical
generators is not completely clear (in the sense of what exactly is held fixed on the boundary).
The problem is also that in other articles \cite{koga}, \cite{hotta&sasaki} it is shown
that the central charge should be zero. Due to this discrepancies we want to analyze again this problem 
starting this time directly with the non-rotating case, which seems to be more difficult, and paying special attention
to the different possible boundary conditions on the horizon and the associated boundary terms of the generators.
We will see that in order to have a boundary term in the Hamiltonian we need to fix the surface gravity on the horizon rather than the metric.
This is because with fixed bolt metric we will see there is a bolt term in the action which
is canceled by the Legendre transform and so absent in the Hamiltonian and therefore the canonical generators acquire no boundary term associated to the bolt. 
The crucial point in this calculations in fact is   that in order to find the central term of (\ref{poisson})
one uses the fact that the bulk part of the generators is a sum of constraints and therefore zero on shell.
On shell therefore the Poisson algebra (\ref{poisson}) reduces to the Dirac algebra of the boundary terms.
\begin{equation}
\left\{ J[\xi ] , J[\eta ]    \right\} _D = J\left[ [\xi , \eta ]_{SD}  \right] +K(\xi , \eta )  \; . \label{dirac} 
\end{equation}
Without boundary terms the generator algebra reduces to the constraint algebra and does not admit central extensions.
Fixing the surface gravity on the bifurcation the generators acquire a boundary term associated to the bolt, but unfortunately
we will see that also in this case there will be no central extension of the generator algebra because of the falloff conditions
we have to impose on the diffeomorphism parameters, at least if we want that the diffeomorphism parameters
and its derivatives have a well defined limit  on the horizon.
While similar results have been found for 2-D dilatonic black holes in \cite{vanzo} in this paper the four-dimensional case is considered.

\section{Boundary terms of the action and boundary conditions}
In order to have a well defined least action principle, it is necessary to have a differentiable action.
This means especially that its variation should consist only of a bulk term without boundary terms. Boundary terms in the 
action arise due to partial integration, where one transforms total divergences in boundary integrals.
A variational principle must also be accompanied by boundary conditions. The most usual boundary condition is to keep the 
variation of the field fixed on the boundary.\\
In the case of the Einstein-Hilbert action
\begin{equation}
I_{EH} = \frac{1}{16\pi } \int _M \sqrt{-g}\, R   \; ,             \label{einsteinhilbert}
\end{equation}
The scalar curvature  contains second derivatives of the metric  and therefore the boundary term contains
variations of the normal derivatives of the metric. So the action (\ref{einsteinhilbert}) with standard boundary conditions is not 
differentiable. Therefore in order to have a well defined variation principle we must add to this action a boundary term that 
cancels the boundary term arising from the variation. The correct action for smooth boundaries is \cite{hawking&gibbons}
\begin{equation}
I = \frac{1}{16\pi } \int _M \sqrt{-g} \, R + \frac{1}{8\pi} \int _{\partial M} K \,  \sqrt{h}   \; ,     \label{einsthilboundary} 
\end{equation}
where $K$ is the extrinsic curvature of the boundary and $h$ is the determinant of the boundary metric.
When calculating the Hamiltonian from the action we have to keep this boundary terms. The consequence is that also
the Hamiltonian has boundary terms. The boundary terms of the Hamiltonian can also be found directly making the 
variation of the bulk term of it as done for asymptotically flat spaces in \cite{regge&teitelboim}.\\
In the canonical formalism we have spacetime regions of the form $ M = [t_1 , t_2 ] \times  \Sigma$, where $\Sigma$ is a spacelike 
hypersurface. For such a region the boundary has the form
\begin{equation}
\partial M = \Sigma _1 \cup \Sigma _2 \cup B^3 \; ,           \label{canboundary}
\end{equation} 
where $\Sigma _{1,2}$ are the initial and final hypersurfaces and $B^3$ is the timelike  3-boundary spatially bounding the system.
Using this notation and with this kind of boundary  with this kind of boundary the action (\ref{einsteinhilbert}) becomes,
calling $\Theta$ the extrinsic curvature of $B^3$ and $m$ the determinant of it's metric
\begin{equation}
I_0 = \frac{1}{16\pi} \int _M \sqrt{-g} \, R - \frac{1}{8\pi} \int ^{\Sigma _2} _{\Sigma _1} K \, \sqrt{h} +\frac{1}{8\pi} \int _{B^3} \Theta \, \sqrt{m} \; ,
                                                                                           \label{boundaryaction}
\end{equation}
where the integral $\int ^{\Sigma _2} _{\Sigma _1}$  means the integral over $\Sigma _2 $ minus the integral over $\Sigma _1$.
To be precise the action (\ref{einsthilboundary}) is correct for smooth boundaries.  With a boundary of the form (\ref{canboundary})
the intersections of $B^3$ with $\Sigma _{1,2}$ are non-smooth. The variation of the action acquires also boundary terms for the joints \cite{hayward}.
Performing in fact the variation of the actions without boundary terms   introducing the notations 
\begin{equation}
v_{ab} \equiv \delta g_{ab} \; \; \; ; \; \; \; v \equiv g^{ab} v_{ab}   \; ,
\end{equation}
we obtain the formula \cite{wald}
\begin{equation}
\delta I_{EH} = \mathrm{bulk \, terms} +\frac{1}{16\pi} \nabla _a \left( -\nabla _b v^{ab} +\nabla ^a v\right) \equiv \mathrm{bulk \, terms}
 +\frac{1}{16\pi} \nabla_a \delta \mathcal{Z} ^a  \; .
\end{equation}
The total divergence comes from the variation $g^{ab}\delta R_{ab}$. The total divergence can be converted in a boundary integral and so discarding the 
bulk terms giving the equations of motion we can write
\begin{equation}
\delta I_{EH} = \frac{1}{16\pi} \int _{\partial M} n_a \delta\mathcal{Z} ^a \; .   \label{bulvariation}
\end{equation}
Focusing for a moment on the final and initial spacelike hypersurfaces $\Sigma _{1,2}$ we can write  $\delta \mathcal{Z} ^a$ in terms of 
$K_{ab}$ and $h_{ab}$ 
\begin{equation}
u_c \delta \mathcal{Z} ^c = -2\delta K - K^{ab} \delta h_{ab} + D _a \delta u^a \; .      \label{versori0}
\end{equation} 
Considering now the variation of the boundary terms associated to $\Sigma _{1,2}$ in the action $I_0$ (\ref{boundaryaction}) and combine them
with the boundary terms of the variation of the bulk action we obtain
\begin{equation}
-\frac{1}{16\pi} \int ^{\Sigma _2} _{\Sigma _1} \sqrt{h} u_c \delta \mathcal{Z} ^c - \frac{1}{8\pi} \delta \int ^{\Sigma _2} _{\Sigma _1}\sqrt{h} K
= \int _{\Sigma _1} ^{\Sigma _2} P^{ab} \delta h _{ab} -\frac{1}{16\pi}\int ^{B_2} _{B_1} \tilde{\xi} _a \delta u^a \sqrt{\sigma}
                                     \label{pezzo1}
\end{equation}
The boundaries $B_{1,2}$ are the intersections of the hypersurfaces $\Sigma _{1,2} $ with the 3-boundary $B^3$. 
The vector $\tilde{\xi}$ is the normal to $ B_{1,2}$ as considered embedded in $\Sigma_{1,2}$,  it is  equal to the 
normal $\xi$ of $B^3$ only if the boundaries are orthogonal.\\
The procedure for the 3-boundary terms of  $B^3$ is analogous as made before for the initial and final hypersurfaces. Again we have 
\begin{equation}
\xi _a \delta \mathcal{Z} ^a = -2\delta \Theta - \Theta ^{ab} \delta m_{ab} +\tilde{D} _a \delta \xi ^a  \; .
\end{equation} 
Again we put together the variation of the boundary terms of $I_0$ and the boundary terms of the bulk variation
\[
-\frac{1}{16\pi} \int _{B^3}  \sqrt{-m} \xi_c \delta \mathcal{Z} ^c - \frac{1}{8\pi} \delta \int  _{B^3}\sqrt{-m} \Theta
\]
\begin{equation}
=- \int _{B^3} \Pi^{ab} \delta m _{ab} -\frac{1}{16\pi}\int ^{B_2} _{B_1} \tilde{u} _a \delta \xi^a \sqrt{\sigma}
                                                                      \label{pezzo2} 
\end{equation}
Here the vector $\tilde{u}$ is the normal to $B_{1,2}$ as considered embedded in $B^3$ and again it is equal to the normal $u$ only when the 
boundaries are orthogonal. \\
Now we want to put the joint pieces containing $\tilde{u} _a \delta \xi ^a$ and $\tilde{\xi} _a \delta u^a$ together. To do this
we notice that the vectors $\tilde{u}$ and $\tilde{\xi}$ can be written as
\begin{equation}
\tilde{u} = \lambda ( u -\eta\xi ) \; \; \; ; \; \; \; \tilde{\xi} = \lambda ( \xi + \eta u )  \; ,    \label{versori1} 
\end{equation}
where $\eta$ is the scalar product $\eta \equiv u\cdot \xi$ normalization factor $\lambda$ is 
\begin{equation}
\lambda = (1 + \eta ^2 ) ^{-\frac{1}{2}} \; .                                                          \label{versori2}
\end{equation}
In order to put together the terms with  $\tilde{u} _a \delta \xi ^a$ and $\tilde{\xi} _a \delta u^a$ we can introduce the boost parameter $\theta$
defined as
\begin{equation}
\sinh \theta = u \cdot \xi \equiv \eta \; .          \label{boostparameter}
\end{equation}
Noticing now that $u\delta u = 0 $ we can write
\begin{equation}
\tilde{\xi} \delta u = \lambda \xi \delta u = \lambda \delta \eta = \delta \theta           \label{pezzo3}
\end{equation} 
we can therefore write
\begin{equation}
\int ^{B_2} _{B_1} \tilde{\xi} _a \delta u^a \sqrt{\sigma} = \int^ {B_2} _{B_1} \delta \theta \sqrt{\sigma} \; .   \label{pezzo4}
\end{equation}
In the same way we compute the term coming from the 3-boundary
\begin{equation}
\tilde{u} \delta \xi = \lambda u\delta\xi = \lambda \delta \eta = \delta \theta          \label{pezzo6}                    
\end{equation}
and therefore gives the same contribution as the term coming from the spacelike boundary (\ref{pezzo3}). We can write
\begin{equation}
\int ^{B_2} _{B_1} \tilde{u} _a \delta \xi ^a \sqrt{\sigma} = \int^ {B_2} _{B_1} \delta \theta \sqrt{\sigma} \; .              \label{pezzo5}
\end{equation}
Putting now together the single pieces of the variation (\ref{pezzo1} , \ref{pezzo2} ) and using (\ref{pezzo4} , \ref{pezzo5} ) we obtain 
eventually for the complete variation of the action the expression
\[
\delta I_0 = \frac{1}{16\pi} \int _M G_{ij} \delta g^{ij} + \int ^{\Sigma _2} _{\Sigma_1} P^{ab} \delta h_{ab} \sqrt{h}
\]
\begin{equation} 
- \int _{B^3} \Pi ^{ab} \delta m_{ab} \sqrt{-m} - \frac{1}{8\pi} \int ^{B_2} _{B_1} \sqrt{\sigma} \delta \theta  \; ,      \label{varbolt}  
\end{equation}  
 Let us now analyze the terms on the
right side of (\ref{varbolt}). The first term gives the equations of motion the second and third are terms are linear in  to the variation
of the boundary metric and therefore with our boundary conditions zero. The last term in general is not zero.
Therefore in the presence of nonsmooth intersections of boundaries the correct action for fixed boundary metric is 
\begin{equation}
I' = \frac{1}{16\pi}\int _M \sqrt{-g} \, R - \frac{1}{8\pi} \int ^{\Sigma _2} _{\Sigma _1} K \, \sqrt{h} +\frac{1}{8\pi} \int _{B^3} \Theta \, \sqrt{m}
+\frac{1}{8\pi}\int ^{B_2} _{B_1} \sqrt{\sigma} \theta  \; .                 \label{boltaction} 
\end{equation}
The last term in the action, i.e. the joint term in literature is also called ``tilting term'' \cite{hawking&hunter}. It is zero in the
case, that the hypersurfaces $\Sigma$ are orthogonal . We have considered up to now non-smooth boundaries given by the intersection of $\Sigma$ with $B^3$. 
Another case of non-smooth boundary can be given by two intersecting spacelike hypersurfaces.\\
Now let us consider a static black hole. It's timelike killing vector is null on the horizon, this means that in the standard foliation $t= \mathrm{const}$ 
all the spacelike hypersurfaces intersect in a 2-D sphere called the bifurcation. Therefore in this situation the action describing a spacetime containing 
a black hole has a non-smooth boundary in the bifurcation given by two
intersecting spacelike hypersurfaces $\Sigma _{1,2}$. This kind of joint in literature is  also called ``bolt'' \cite{hawking&gibbons}. 
In the case of a ``bolt'' all the computation done before leading to (\ref{varbolt}) can be repeated. Using in fact again  (\ref{versori0})
we obtain now as contribution from the two spacelike hypersurfaces, converting the total divergences in an integral on the joint, the joint contribution
\begin{equation}
\Delta = \frac{1}{16\pi} \int _B  \sqrt{\sigma} \left( \tilde{\xi} _2 \cdot \delta u_1 - \tilde{\xi}_1 \cdot \delta u_2   \right)        \; .
\end{equation}  
The vector $\tilde{\xi}_2$ is the normal to the bolt as considered embedded in $\Sigma ^1$ and the vector $\tilde{\xi}_1$ is the normal to the 
bolt as considered embedded in $\Sigma _2$. Now again as in (\ref{versori1} ,\ref {versori2}) we can write the vectors $\tilde{\xi}_1$ and $\tilde{\xi}_2$
as linear combination of  $u_1$ and $u_2$ with the only change that now scalar product is $\eta \equiv u_1 \cdot u_2$.
The boost parameter $\theta$ this time is  defined as 
\begin{equation}
\cosh \theta = -  u_1 \cdot u_2     \label{boostparameter2}  \; .
\end{equation}
This is because in the bolt case, being the intersecting hypersurfaces both spacelike,  their normals cannot be orthogonal.
Following now the same procedure as in (\ref{pezzo3} , \ref{pezzo6}) it is immediate to proof that
\begin{equation}
\tilde{\xi}_2 \cdot \delta  u_1 - \tilde{\xi}_1 \cdot \delta u_2 = -2\delta \theta  \; .
\end{equation}
The total bolt contribution from the bulk variation is therefore 
\begin{equation}
\Delta = -\frac{1}{8\pi} \int _B \sqrt{\sigma} \delta \theta
\end{equation}
Therefore if we treat the event horizon as a boundary and using the standard foliation the correct action by fixed bolt metric is   
\begin{equation}
I= I_0 +  \frac{1}{8\pi}\int _{B} \sqrt{\sigma} \theta  \; . \label{bifurcationaction}
\end{equation}
Let us now notice that for a Kerr black hole the killing vector $\partial _t$ on the horizon goes to zero only in two points, namely
the ``north pole'' and the ``south pole''. Having only two points in which the $t= \mathrm{const}$ hypersurfaces intersect there is no boundary term for this 
intersection. The action therefore acquires no extra term at least in the standard foliation. Having in this case only 2 points as intersection
also the Hamiltonian won't have boundary terms associated to the horizon. Therefore the technique to find a central extension of the boundary terms 
Dirac algebra (\ref{dirac}) like in \cite{brown&henneaux} seems to work only for the non-rotating case, at least in the standard foliation.\\
Now we have seen that the boundary term of the action associated to the bifurcation (\ref{bifurcationaction}) works with boundary condition of 
fixed boundary metric. This is the most usual boundary condition but surely not the only possible. One for example can also fix the normal derivative 
of the  boundary metric. In the case of the bifurcation this means to fix the surface gravity.\\ 
Let us now analyze what boundary term we have to associate to the bifurcation in the case that we keep the surface gravity fixed instead of the
metric. In order to do this let us write the parameter $\theta$ of (\ref{boostparameter} for the case of a non-rotating black hole.
In this case we have metric of the form
\begin{equation}
ds^2 = -N^2 dt^2 + N^{-2} dr^2 + r^2 d\Omega ^2   \; . \label{nearhorizon}
\end{equation}
In the bifurcation all the constant $t$ hypersurfaces intersect and so the normal to the hypersurfaces there is not well defined.
to compute the scalar product in (\ref{boostparameter2}) we can parallel transport the normal of one hypersurface, say $\Sigma _{t_1}$ to another
say $\Sigma _{t_2}$ along an $r=\mathrm{const}$ curve.
Using (\ref{nearhorizon}) the only nontrivial parallel transport equations become
\begin{equation}
\dot{u}^t + \Gamma ^{t}_{tr} \dot{t} u^r =0 \; \; \; ; \; \; \; \dot{u}^r +\Gamma ^{r}_{tt} \dot{t} u^t + \Gamma ^{r}_{tr} \dot{t} u^r =0
\end{equation}
with 
\begin{equation}
\Gamma ^{t }_{tr} = \frac{1}{2} N ^{-2}  (N ^2 ) ' \; \; ; \; \;  \Gamma ^{r }_{tt} =\frac{1}{2} N ^2  (N ^2 ) ' \; \; ; \; \;  \Gamma ^{r}_{tr} =0  \; .
\end{equation}
he solution therefore is 
\begin{equation}
u^t = N^{-1} \cosh (\kappa t) \; \; \; ; \; \; \; u^r = -N \sinh (\kappa t)  \; .
\end{equation}
The scalar product of the normals of $\Sigma _1$ and $\Sigma _2$ is then 
\begin{equation}
u_1 \cdot u_2 = - \cosh (\kappa \Delta t)  \; .
\end{equation}
the boost parameter $\theta$ is then
\begin{equation}
\theta = \kappa \Delta t 
\end{equation}
this means that the bifurcation term of the action can be written as 
\begin{equation}
\frac{1}{8\pi} \int _{B} \kappa dA dt \; .       \label{boltcanonical} 
\end{equation}
Now let us remember that the origin of the bolt term is to cancel the term linear in $\delta \theta$ in (\ref{varbolt}). Now being $\theta$
proportional to the surface gravity in the case we keep the surface gravity fixed in the variation the $\delta \theta$ term is then zero.
We can therefore conclude that in the case we keep the surface gravity fixed there is no bifurcation term in the action.

\section{ Boundary terms of the Hamiltonian}
We have up to now seen that the origin of a possible boundary term associated the horizon depends on the foliation and on the boundary conditions.
Having found the boundary terms for the action let us see how they reflect in the Hamiltonian and the canonical generators.
Now in order to find the boundary terms of the canonical generators let us write the action without tilting term (\ref{boundaryaction}) in canonical form
expressing everything in function of the canonical momenta and the hypersurface 3-metric $h_{ab}$. The result is \cite{hawking&hunter} 
\[
I_0 = \int _M \left( P^{ab} \dot{h} _{ab} - N\mathcal{H} - N^i \mathcal{H} _i \right) d^4 x -2 \int dt \int _{B_t} h^{-1/2}P^{ab} N_a \tilde{\xi} _b
\sqrt{\sigma} d^2x 
\]
\begin{equation}
- \frac{1}{8\pi} \int _M \nabla _a Z^a \sqrt{-g} d^4 x -\frac{1}{8\pi} \int _{\Sigma _1} ^{\Sigma _2} K \sqrt{h} d^3 x + \frac{1}{8\pi}
\int _{B_3} \Theta \sqrt{-m} dt d^2 x      \; .              \label{hahu}
\end{equation}
The functions $\mathcal{H}$ and $\mathcal{H}_i$ are the Hamiltonian constraints.
The boundary $B_t$ is the foliation of $B_3$ in the form
\begin{equation}
B_t = B_3 \cap \Sigma _t
\end{equation} 
The term $Z^a$ is given by
\begin{equation}
Z^a =\nabla _u u^a - u^a \nabla _b u^b
\end{equation}
and so we have that
\begin{equation}
Z^a u_a =K          \label{formula}
\end{equation}
Now as next we must convert to a boundary integral  the $\nabla _aZ^a$ term.
Notice that due to (\ref{formula}) the integral $\int _{\Sigma _1} ^{\Sigma _2}$ is therefore 
canceled. It survives only the boundary integral over $B_3$. Now in order to be able to read out the Hamiltonian from (\ref{hahu})
we have  to factorize out a $\int dt$ term
from the boundary integrals. To do this let us notice that $B_3$ is foliated by $B_t$ and therefore we can write
\begin{equation}
\sqrt{-m} = N \lambda \sqrt{\sigma} \; ,
\end{equation} 
where $\sigma _{ab}$ is the metric induced on each $B_t$ by $m_{ab}$ and $h _{ab}$ and $\lambda$ is defined as $\lambda = \cosh \theta$.
The surviving boundary terms can therefore be factorized with $\int dt$
\[
\frac{1}{8\pi} \int _{B^3} \Theta \sqrt{-m} dtd^2 x -\frac{1}{8\pi} \int _{B^3} \xi _a Z^a \sqrt{-m} dt d^2 x 
\]
\begin{equation}
= \frac{1}{8\pi} \int dt \int _{B_t} N \tilde{\Theta} \sqrt{\sigma} d^2 x - \frac{1}{8\pi} \int dt \int _{B_t} N \lambda \tilde{u} ^a \partial _a (\theta)
\sqrt{\sigma} d^2 x  \; ,            \label{factorization}
\end{equation}
where $\tilde{\Theta} $ is the trace of the extrinsic curvature of $B_t$ as embedded in $\Sigma _t$ and $\tilde{u}$ is
the normal to $B_t$ as considered embedded in $B^3$. If we choose the time flow $\tau$ to be tangent to $B^3$ we can write
\begin{equation}
\tau ^a = N \lambda \tilde{u} ^a 
\end{equation}
and therefore
\begin{equation}
N\lambda \tilde{u} ^a \partial _a \theta = \dot{\theta} \; .
\end{equation}
Therefore the last term in (\ref{factorization}) becomes
\begin{equation}
-\frac{1}{8\pi} \int _{B}  \theta \sqrt{\sigma} d^2 x + \frac{1}{8\pi} \int dt  \int _{B_t} \theta \dot{\sqrt{\sigma}} d^2 x      \; .
                                                                            \label{cancellation} 
\end{equation} 
The first term cancels the tilting term in (\ref{boltaction}) whereas the second term in the case that the joint is a bifurcation is zero because
it is static per definition.\\
The Hamiltonian resulting from (\ref{hahu}) is using (\ref{factorization} , \ref{cancellation}) 
\begin{equation}
H = \int _{\Sigma _t} d^3 x \sqrt{h} \left( N\mathcal{H} + N^i \mathcal{H}_i \right) - \frac{1}{8\pi} \int_{B_t} 
\left( N \tilde{\Theta} - 16\pi h^{-1/2} P^{ab} N_a \tilde{\xi}_b   \right)  \sqrt{\sigma}  
\end{equation}
We see so that there is no contribution from the bifurcation to the Hamiltonian boundary terms. 
The action (\ref{boltaction}) was the correct one in the case that the boundary metric was kept fixed. If we now take the case of a black hole 
bifurcation as joint with the surface gravity held fixed instead of the bolt metric we have seen that there is no tilting term contribution to the action.
Therefore there is then a tilting term in the Hamiltonian that comes from (\ref{cancellation}) that is now not canceled from the action.
The Hamiltonian for fixed surface gravity $H'$ is therefore
\begin{equation}
H' = H - \frac{1}{8\pi} \int _{B} \theta \sqrt{\sigma} d^2x
\end{equation}
In this case there is a boundary term contribution from the bifurcation to the Hamiltonian. 
We have in the case of the bifurcation therefore the situation, that fixing the metric there is a tilting term in the action but not in the 
Hamiltonian. Whereas fixing the surface gravity there is a no tilting term in the action but there is one in the Hamiltonian. 
Therefore if we want to have a on shell a Dirac algebra for the black hole we necessarily must fix the surface gravity.
This is physically reasonable because the surface gravity gives the temperature of the black hole and our calculations attempt 
precisely to describe the black hole thermodynamics.\\
By fixed surface gravity therefore using (\ref{boltcanonical}) the bolt term associated to the canonical generator of the vector field
$\xi$ is
\begin{equation}
J[\xi ] = - \frac{1}{8\pi} \int _{B} n^c D_c \xi ^{\bot}   \label{canonicalbolt} 
\end{equation}

\section{Fall off conditions and Poisson algebra}

Now we want to study the deformations of the $r-t$ plane that preserve the surface gravity of the horizon.
In order to do this we must find the falloff condition of the vector fields generating the diffeomorphisms. Let us now find the most general expression
for a near horizon metric. As explained before we must start from the non-rotating case in order to have a bifurcation in the standard foliation.
Making the Ansatz of spherical symmetry we have
\begin{equation}
ds^2 = -N^2 (r,t) + A^2 (r)dr^2 +r^2d \Omega ^2
\end{equation}
We have to impose some conditions on the functions $N$ and $A$.\\
First of all we notice that the existence of a bifurcation implies the vanishing of the lapse function $N$ on the horizon
\begin{equation}
N^2 (r_{+},t) =0   \; .
\end{equation}
We are dealing with a system of fixed surface gravity, where the surface gravity is defined as 
\begin{equation}
\lim_{r \rightarrow r_{+}} \frac{\partial_r N}{A} = \kappa
\end{equation}
We must also impose the topology of the black hole in order to distinguish it from flat spacetime. In the Euclidean case the black hole has the topology
$R^2 \times S^2 $ and therefore the Euler characteristic is $\chi = \chi (\mathrm{disk} ) \times \chi (\mathrm{sphere})$ \cite{brownetal} and being the Euler 
characteristic of the sphere $2$, the Euler characteristic of the black hole is  
\begin{equation}
\chi = 2   \; , 
\end{equation}
whereas the flat spacetime has $\chi =0$. Calculating $\chi$ we obtain
\begin{equation}
\chi = 2 \left( 1-A^{-1} (r_{+}) \right)   \; .
\end{equation}
We obtain so the condition on $A$
\begin{equation}
A^{-1}(r_{+}) =0
\end{equation}
If we impose the Hamiltonian constraints we obtain the form for$A^2$
\begin{equation}
A^2 = \left(1- \frac{r_{+}}{r} \right) ^{-1}
\end{equation}
Using this conditions we obtain the explicit form for the lapse
\begin{equation}
N^2 = 4 \kappa^2 r_{+} (r-r_{+}) + a(t)(r-r_{+})^2
\end{equation}
Now on shell we have $r_{+}=\frac{1}{2 \kappa}$
therefore putting all together our diffeomorphisms must preserve the following conditions
\begin{equation}
N^2 =g_{tt}= 2 \kappa (r-r_{+}) + \mathcal{O}(r-r_{+})^2    \label{condition1}
\end{equation}
and for $g^{rr}$ we have then
\begin{equation}
g^{rr} = N^2 + \mathcal{O} (N^3 )       \label{condition2}
\end{equation} 
We must now search for vector fields which preserve the two conditions (\ref{condition1} ) and (\ref{condition2}). Satisfying this conditions the
diffeomorphisms automatically preserve the surface gravity. For a vector field $\xi$ in the $r-t$ plane the variation of $g_{tt} $ is given by
\begin{equation}
\delta _{\xi} g_{tt} = \mathcal{L} _{\xi} g_{tt} = \partial _r g_{tt} \xi^r + 2g_{tt}\partial _t \xi^t = \mathcal{O} (N^3 )
\end{equation}
and therefore
\begin{equation}
\xi ^r = - \frac{N^2}{\kappa} \partial _t \xi^t + \mathcal{O}(N^3)        \label{radialdiff}
\end{equation}
The variation of the $g^{rr}$ component is now
\begin{equation}
\delta _{\xi} g^{rr} = \partial _r g^{rr} \xi ^r + 2g^{rr} \partial _r \xi ^r = \mathcal{O} (N^3)
\end{equation}
The boundary term of the generator   (\ref{canonicalbolt}) implies only the $\xi ^t$ component and therefore let us use (\ref{radialdiff}) 
in the last variation in order to find the form of $\xi ^t$
\begin{equation}
=- 2N ^2 \dot{\xi} ^t -4 N^2 \dot{\xi} ^t - \frac{2N^4}{\kappa} (\dot{\xi} ^{t})' =\mathcal{O} (N^3)
\end{equation}
This means that 
\begin{equation}
\xi ^t = \mathcal{O} (1)  \; \; ; \; \;  \xi^r = \mathcal{O}(N^3)  \; \; ; \; \;\xi ^{\bot} =\mathcal{O}(N) .
\end{equation}
Notice that this falloff conditions ensure also the vanishing of the extra diagonal components of the metric on the horizon
\begin{equation}
\delta g_{tr} = \mathcal{O} (N)  \; .
\end{equation} 
Therefore with our boundary conditions we have a boundary term in the generators and therefore there is the possibility
of the existence of a nontrivial central charge.
Now we are studying diffeomorphisms that act near the horizon and therefore it is reasonable to impose that the vector fields defining the
diffeomorphisms and its derivatives  should have a well defined limit on the horizon 
We must therefore compute the Poisson brackets for two generators.
This calculation can be done  using the results in \cite{brown&lau&york} where the variation of the Hamiltonian
under a quasilocal boost is done and inserting our falloff
conditions. The boundary terms coming from this variation can be computed going on shell obtaining eventually for
our boundary conditions
\begin{equation}
\left\{ H[\xi ] , H[\eta ]    \right\}_{PB} = H[\xi , \eta ] _{SD} \; ,
\end{equation} 
where $H$ is the generator with the correct boundary term. We conclude therefore that there is no central extension of the algebra in 
this case also if the boundary terms are nonzero.\\
As said before this is true at least if we want the vector fields and its derivatives to have well defined limit on the horizon. 
\section{Conclusions}
We have analyzed two cases of boundary conditions: the case of fixed bolt metric and the case of fixed surface gravity.
In the first case there were no boundary terms associated to the
bolt in the canonical generators and so becoming the generator algebra the constraint algebra it does not admit a central extension.
In the second case there were nonzero boundary terms associated to the bifurcation, but the computation of the Poisson brackets shows that there is no
central extension in the generator algebra.
The calculation of Strominger \cite{strominger} seems therefore limited to the $BTZ$ black hole. The $BTZ$ black hole is in fact 
embedded in an $AdS$ spacetime. The fact that one finds a nonzero central charge 
for the generators, that preserve the $AdS$ structure at infinity becomes a particular case of the  $AdS/CFT$ correspondence \cite{maldacena}.
In this case the relevant CFT at spatial infinity is the Liouville theory \cite{henneauxL}. The quantum version of this Liouville theory living on the $AdS$
boundary is described in \cite{chen}, where  again the BTZ entropy is obtained.
There is also a $dS /CFT$ correspondence \cite{strominger2},
\cite{park3}, \cite{park4}, \cite{klemm1}, \cite{klemm2}, \cite{klemm3} but we want to describe a black hole
independently of its embedding.\\
The things may change if  we analyze other physically meaningful boundary conditions. The generalization to other 
boundary conditions is nontrivial and may be object of further investigation in future.   
Also if, with our two boundary conditions which are the most natural,   we found a negative result  this does not mean that it is impossible to count the black hole microstates by means of the Cardy 
formula using a classical central charge.
This negative result in fact only means that the origin of a classical central central charge for a theory describing the black hole is different. 
In fact in the already cited previous works of the author it was shown that  in a dimensionally reduced approach, near horizon, the
black hole dynamics is described by a Liouville theory which has a classical central charge. But in this case the existence of the classical central charge 
is not due to boundary terms in the generators but to the affine scalar behavior of the Liouville field.
 
\section*{Acknowledgements}
The author wants to thank Prof. Luciano Vanzo for his many explanations and suggestions on the subject.
The author thanks also Dr. Valter Moretti and Dr. Nicola Pinamonti for the useful discussions.


\begin{thebibliography}{99}


\bibitem{hawking&bardeen&carter} Bardeen , Carter and  Hawking, The Four Laws of Black Hole  Mechanics.
{\it Comm. Math. Phys.} 31; 161 (1973) 


\bibitem{hawking}   Hawking S.W. , Particle Creation by Black Holes.
{\it Comm. Math. Phys.} 43, 199 (1975)


\bibitem{bekenstein} Bekenstein J. D. , Black Holes and Entropy.
{\it Phys. Rev.} D7 , 2333,  (1973)

\bibitem{bekenstein2} Bekenstein J. D., Black holes and the Second Law.
{\it Lett. Nouvo Cim.} 4, 737 (1972)


\bibitem{vafa}  Strominger A. and Vafa C., Microscopic Origin of the Bekenstein-Hawking Entropy.
{\it Phys. Lett.} B379, 99 (1996)

\bibitem{ashtekar} Ashtekar A., Baez J., Corichi A., Krasnov K., Quantum Geometry and Black Hole Entropy.
{\it Phys. Rev.Lett. } 80, 904 (1998)

\bibitem{sakharov}  Frolov V.P. and Fursaev D.V., Mechanism of Generation of Black Hole Entropy in Sakharov's Induced Gravity.
{\it Phys. Rev} D56,  2212 (1997)       


\bibitem{cardy} Cardy J. A. , Operator Content of Two Dimensional Conformally Invariant Theories.
{\it Nuc. Phys. } D270 , 186 (1986)

\bibitem{arnold}    Arnold V. I. , Mathematical Methods of Classical Mechanics. Berlin, Heidelberg, New York: Springer (1978)


\bibitem{solodukhin} Solodukhin S. N., Conformal description of Horizon's States.
{\it Phys. Lett.} B454, 213 (1999)

\bibitem{brustein}  Brustein R., Causal Boundary Entropy from Horizon Conformal Field Theory.
{\it Phys. Rev. Lett} 86, 576 (2001)

\bibitem{giacomini&pinamonti} Giacomini A. and Pinamonti N. .  Black Hole Entropy from Classical Liouville Theory.
{\it JHEP} 02 (2003) 014

\bibitem{giacomini2}  Giacomini A. . Two Dimensional Conformal Symmetry and the Microscopic Interpretation of Black Hole 
Entropy (Ph.D. thesis). hep-th 0403183

\bibitem{giacomini3} Giacomini A. . Classical Liouville Theory and the Microscopic Interpretation of Black Hole Entropy
(To be published in the proceedings of the 5. Int. Conference on Symmetry in Nonlinear  Mathematical Physics; Kyiv Ukraine 23.6-29.6.2003).
hep-th 0403182

 
\bibitem{brown&henneaux} Brown J. D. and Henneaux M. , Central Charges in the Canonical Realisation of 
Asymptotic Symmetries : An Example from Three Dimensional Gravity .
{\it Comm. Math. Phys.} 104 , 207 (1986)

\bibitem{teitelboim} Teitelboim C. , Commutators of Constraints Reflect the Spacetime Structure.
{\it Ann. Phys.} 79, 542 (1973) 

\bibitem{brown&henneaux2} Brown J. D. and Henneaux M., On the Poisson Brackets of Differentiable Generators in Classical Field Theory.
{\it J. Math. Phys.} 27, 489 (1986) 

\bibitem{strominger}  Strominger A., Black Hole Entropy from Near-Horizon Microstates. {\it JHEP} 9802, 009  (1998)


\bibitem{BTZ} Ba\~nados M., Teitelboim C., Zanelli J., Black Holes in Three-Dimensional Spacetime.
{\it Phys. Rew. Lett. } 69, 1849 (1992)

\bibitem{cadoni1} Cadoni M., Mignemi S., Asymptotic Symmetries of $AdS_2$ and Conformal Group in $d=1$.
{\it Nucl. Phys. } B557, 165 (1999)

\bibitem{cadoni2} Cadoni M., Cavagli\`a M., Two Dimensional Black Holes as Open Strings: A New Realization of the $AdS / CFT$ Correspondence.
{\it Phys. Lett.} B499, 315 (2001) 

\bibitem{hawking&gibbons}  Gibbons G. W. and Hawking S. W., Action Integrals and Partition Functions in Quantum Gravity.
{\it Phys. Rev.} D15, 2752 (1977)

\bibitem{regge&teitelboim} Regge T. Teitelboim C., Role of Surface Integrals in the Hamiltonian Formulation of General Relativity.
{\it Ann. Phys.} 88, 286 (1974)

\bibitem{hayward} Hayward G., Gravitational Action for Spacetimes with Nonsmooth Boundaries.
{\it Phys. Rev.} D47, 3275 (1993)


\bibitem{wald} Wald R. M. , General Relativity . The University of Chicago Press (1984)

\bibitem{hawking&hunter} Hawking S. W. and Hunter C. J., The Gravitational Hamiltonian in the Presence of Non-Orthagonal Boundaries.
{\it Class. Quant Grav.} 13, 2735 (1996) 

\bibitem{park} Park M. and ho J. , Comments on `` Black Hole Entropy from Conformal field Theory in Any Dimension'' .
{\it Phys. Rev. Lett.} 83 , 5595 (1999)

\bibitem{soloviev} Soloview V. O. Black Hole Entropy from Poisson Brackets.
{\it Phys. Rev.} D61  , 27502 (2000)


\bibitem{carlip} Carlip S. , Black Hole Entropy from Conformal Field Theory in Any Dimension .
{\it Phys. Rev. Lett. } 82 , 2828 (1999)


\bibitem{ghosh} Dreyer O. , Ghosh A. and Wisniewski J., Balck Hole Entropy Calculations Based on Symmetries.
{\it Class. Quant. Grav.} 18, 1929 (2001)

\bibitem{carlip2} Carlip S. , Entropy from Conformal Field Theory at Killing Horizons. {\it Class. Quant. Grav} 16 , 3327 (1999)

\bibitem{silva} Silva S., Black Hole Entropy and Thermodynamics from Symmetries
{\it Class. Quant. Grav.} 19, 3947 (2002)

\bibitem{wald1}   Lee J., Wald R. M., Local Symmetries and Constraints. 
{\it J. Math Phys} 31, 725 (1990)

\bibitem{wald2}  Wald R. M., Black Hole Entropy in the Noether Charge. 
{\it Phys. Rev.} D48, 3427 (1993) 

\bibitem{wald3} Iyer V., Wald R. M., Some Properties of Noether Charge and a Proposal for Dynamical Black Hole Entropy.
{\it Phys. Rev.} D50, 846 (1994) 

\bibitem{wald4}  Iyer V. Wald R. M., A Comparison of Noether Charge and Euclidean Methods for Computing the Entropy of Stationary Black Holes.
{\it Phys. Rev.}  D52 4430 (1995)

\bibitem{pallua2}  Cvitan M., Pallua S., Prester P., Higher Curvature Lagrangians, Conformal Symmetry and Microscopic Entropy of Killing Horizons.  
{\it Phys. Lett. } B571, 271 (2003)

\bibitem{park2} Park M. I. Hamiltonian Dynamics of Bounded Spacetime and Black Hole  Entropy: the Canonical Method.
{\it Nucl. Phys.} B634, 339 (2002) 

\bibitem{koga} Koga J. , Asymptotic Symmetries on Killing Horizons. 
{\it Phys. Rev.} D64 , 124012 (2001) 

\bibitem{hotta&sasaki}  Hotta M. , Sasaki K.  and Sasaki T. , Diffeomorphisms on Horizon as an Asymptotic Isometry
of Schwarzschild Black Hole 
{\it Class. Quant. Grav. } 18 , 1823 (2001) 

\bibitem{vanzo} Pinamonti N. and Vanzo L. . Central Charges and Boundary Fields for Two Dimensional  Dilatonic Black holes.
hep-th 0312065

\bibitem{ADM} Arnowitt R. Deser S. Misner C.W., Gravitation an Introduction to modern research. 
edited by L. Witten (John Wiley, New York) p. 227-265

\bibitem{brownetal} Brown J. D., Comer G. L., Martinez E. A., Melmed J., Whiting B. F. , York J. W. Jr., Thermodynamics Ensembles and 
Gravitation
{\it Class. Quant. Grav.} 7, 1433 (1990)

\bibitem{maldacena} Maldacena J., The Large N Limit of Superconformal Field Theories and Supergravity.
{\it Adv. Theor. Math. Phys. } 2, 231 (1998) 

\bibitem{henneauxL} Coussaert O., Henneaux M. and van Driel P., The Asymptotic Dynamics of Three-Dimensional Einstein Gravity with a Negative
Cosmological Constant.
{\it Class. Quant. Grav.} 12, 2961 (1995) 

\bibitem{chen} Chen Y., Quantum Liouville Theory and BTZ Black Hole Entropy.
hep-th/0310234


\bibitem{brown&lau&york}  Brown J. D., Lau S. R., York J. W., Action and Energy of the Gravitational Field.
gr-qc/0010024 
 
\bibitem{strominger2} Strominger A., The dS/CFT Correspondence.
{\it JHEP } 0110 (2001) 034


\bibitem{park3} Park M. I., Statistical Entropy of Three Dimensional Kerr de Sitter Space
{\it Phys. Lett.} B440, 275 (1998)

\bibitem{park4} Park M. I., Symmetry Algebras in Chern Simons Theories with Boundary: Canonical Approach
{\it Nucl. Phys.} B544, 377 (1999) 

\bibitem{klemm1} Klemm D., Some Aspects of the De Sitter CFT Correspondence.
{\it Nucl. Phys.} B625, 295 (2002)

\bibitem{klemm2} Cacciatori S., Klemm D., The Asymptotic Dynamics of de Sitter Gravity in Three Dimensions.
{\it Class. Quant. Grav.} 19, 579 (2002)

\bibitem{klemm3} Klemm D., Vanzo L., De Sitter Gravity and Liouville Theory.
{\it JHEP} 0204, 030 (2002)  




\end{thebibliography}
\end{document}